\documentclass[aps,prl,twocolumn,superscriptaddress,showpacs]{revtex4}

\usepackage{graphics}

\begin{document}


\title{Optimal routing on complex networks}


\author{Bogdan Danila}
\email[]{dbogdan@mail.uh.edu}
\affiliation{Department of Physics, The University of Houston, Houston TX 77004}

\author{Yong Yu}
\affiliation{Department of Physics, The University of Houston, Houston TX 77004}

\author{John A.\ Marsh}
\affiliation{SI International, Beeches Professional Campus, Rome NY 13440}

\author{Kevin E.\ Bassler}
\email[]{bassler@uh.edu}
\affiliation{Department of Physics, The University of Houston, Houston TX 77004}

\date{\today}

\begin{abstract}
We present a novel heuristic algorithm for routing optimization on complex networks. Previously proposed routing optimization algorithms aim at avoiding or reducing link overload. Our algorithm balances traffic on a network by minimizing the maximum node betweenness with as little path lengthening as possible, thus being useful in cases when networks are jamming due to queuing overload. By using the resulting routing table, a network can sustain significantly higher traffic without jamming than in the case of traditional shortest path routing.
\end{abstract}

\pacs{89.75.Hc, 89.20.Hh, 89.75.Da}

\maketitle

\noindent The ever-increasing amount of information and goods transported along complex networks raises the question about their ultimate transport capacity. Generally, the transport capacity of a network is limited by two factors: link capacity (bandwidth in the case of the Internet or wireless networks) and node (router) latency. Traditionally, both fixed and ad-hoc network routing is based on the idea of maintaining a table of the shortest paths (or the best available approximation to the shortest paths) between any two nodes of the network and forwarding the information packages along these paths \cite{Ericsson,Fortz}. The length of a path is computed as the sum of the weights assigned to the links that form the path. In the case of the Internet, link weights are typically assigned manually by operators according to simple rules based on experience \cite{Ericsson}. Recently, a series of heuristic algorithms have also been proposed for network traffic optimization \cite{Ericsson,Fortz,GabrelJOH,AllenJOH,MulyanaETT}. These rules and algorithms are aimed at avoiding or reducing link overload by a judicious link weight assignment. The cost of each link is assessed as a monotonically increasing function of the ratio between traffic and capacity and then the weights are adjusted to minimize the sum of the costs of all links.

This approach, however, does not take into account delays due to node latency (information package processing time) and has the disadvantage that too many of the shortest paths pass through a few nodes, called hubs. As a result, in situations of high network traffic, these hubs will experience congestion (long message queues) and eventually jamming, causing the network to break apart in a multitude of disconnected subnetworks. In light of this behavior, the optimality of the shortest path routing as currently implemented has been recently questioned \cite{YanPRE,Sreenivasan_arXiv,Guimera,EcheniquePRE,EcheniqueEL,ZhaoLaiParkYe,ParkLaiZhaoYe,TB,TKBHK,AshtonPRL}. It has been shown, for example, that dynamic routing protocols which allow for a certain degree of stochasticity or take into account the congestion status of the nearest neighbors significantly improve the transport capacity of a network \cite{EcheniquePRE,EcheniqueEL,ZhaoLaiParkYe,ParkLaiZhaoYe,TB,TKBHK,AshtonPRL}.
A more systematic approach is to find better static (strictly table-driven) routing protocols that avoid the hubs whenever possible and convenient (i.e. when avoiding a hub does not lead to congestion on another node). Recent studies \cite{YanPRE,Sreenivasan_arXiv} have shown that this is possible and propose new routing algorithms which lead to improved transport capacity (quantified by the packet insertion rate at which jamming occurs). An open question is how much larger the actual transport capacity can be than the results presented in Refs.\ \cite{YanPRE,Sreenivasan_arXiv}. In this Letter, we show that significant improvement in the transport capacity of a network can be achieved by systematically adjusting the traffic routing to minimize the maximum betweenness on the network. Our algorithm leads to higher transport capacity than those presented in Refs.\ \cite{YanPRE,Sreenivasan_arXiv}. The transport capacity also exceeds the analytical estimate for its maximum value given in \cite{Sreenivasan_arXiv}. Furthermore, we argue that our algorithm achieves near-optimal routing for uncorrelated scale-free networks.

To facilitate comparison, we use the same network model as in Ref.\ \cite{Sreenivasan_arXiv}. All results presented in this Letter are for undirected, uncorrelated scale-free networks with an exponent of the power-law degree distribution $\gamma=2.5$, generated using the configuration model. The number of nodes $N$ varies between 25 and 1600. For simplicity, we assume that all nodes have the same processing capacity of 1 package per time step and that new information packages are inserted at every node at the same average rate of $r$ packages per time step. The destinations of the packages are chosen at random from among the other $N-1$ nodes on the network. However, the algorithm can be generalized for nodes with different processing capacities and for arbitrary traffic demands. Given a routing table, the betweenness $B_i$ of node $i$ is defined \cite{NewmanPRE} as the sum of all fractional paths that pass through that node. The fraction of times a message passes through node $i$ on its way from a source node $s$ to a target node $t$ is computed as follows: the source node $s$ is assigned a weight 1 and then the weight of every node along each path is split evenly among its predecessors in the routing table on the way from $t$ to $s$ and added to the weights of the predecessors. The average number of packages passing through a given node $i$ is then $<w>_i=r B_i/(N-1)$. Jamming occurs at the critical average insertion rate $r_c$ at which the average number of packages processed by the busiest node reaches unity. Consequently, $r_c$ is given by \cite{Guimera}
\begin{equation}
	r_c=\frac{N-1}{B_{max}},
\end{equation}
\noindent where $B_{max}$ is the highest betweenness of a node on the network. Thus, to achieve optimal routing, the highest betweenness $B_{max}$ should be minimized. An important point is that, even though the minimization procedure pertains to a single scalar quantity, such an optimization algorithm will implicitly reshape the betweenness landscape across the whole network, lowering traffic through the initially busy nodes at the expense of increased traffic through the initially idle nodes until the traffic spreads out and an as narrow as possible betweenness distribution is achieved.
\begin{figure}
	\scalebox{0.32}{\includegraphics{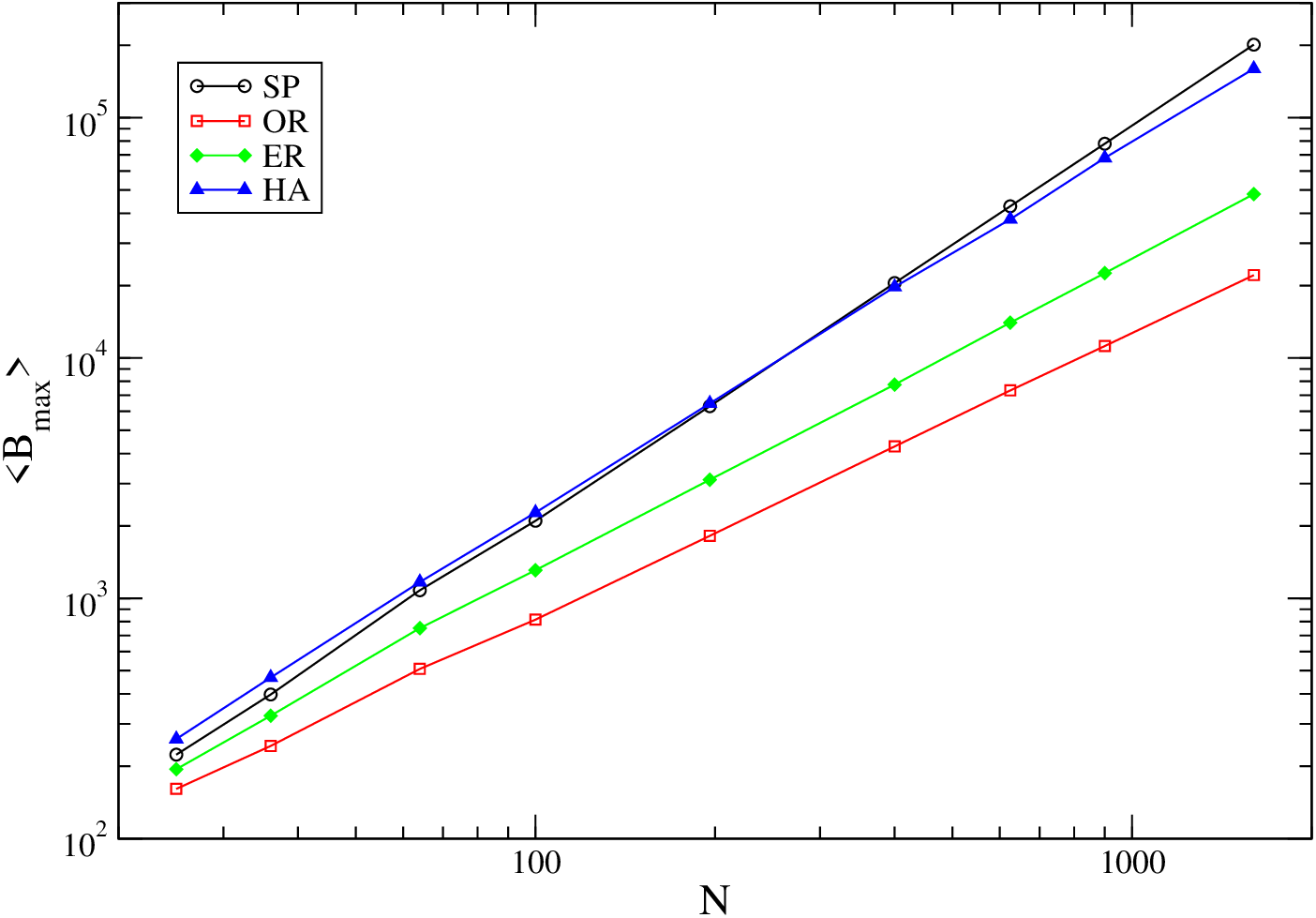}}
	\caption{(Color online) Ensemble average of the network maximum betweenness as a function of the number of nodes for four routing protocols.}
\end{figure}

The problem of finding the exact optimal routing is mathematically tied to the problem of finding the minimal sparsity vertex separator \cite{Sreenivasan_arXiv}, which has been shown \cite{Bui} to be an $NP$-hard problem. This means that the number of flops necessary for the computation of an exact solution increases with the number of nodes $N$ faster than any polynomial. We propose a heuristic algorithm which finds near-optimal solutions for the routing problem in time $\mathcal{O}(N^4)$ at worst ($\mathcal{O}(N^3)$ for one iteration and requiring $\mathcal{O}(N)$ iterations). In its simplest form, the algorithm proceeds as follows:

1. Assign weight 1 to every link and compute the shortest paths between all pairs of nodes.

2. Compute the betweenness of every node.

3. Find the node which has the highest betweenness $B_{max}$ and add 1 to the weight of every link that connects it to other nodes.

4. Recompute the shortest paths. Go back to step 2.

\noindent Note that the algorithm picks the ``least fit" element of a set and changes its parameters. Therefore, it is a from extremal optimization \cite{Boettcher}. However, this algorithm assigns parameters in a deterministic way, unlike many of the other existing extremal optimization algorithms.
\begin{figure}
	\scalebox{0.32}{\includegraphics{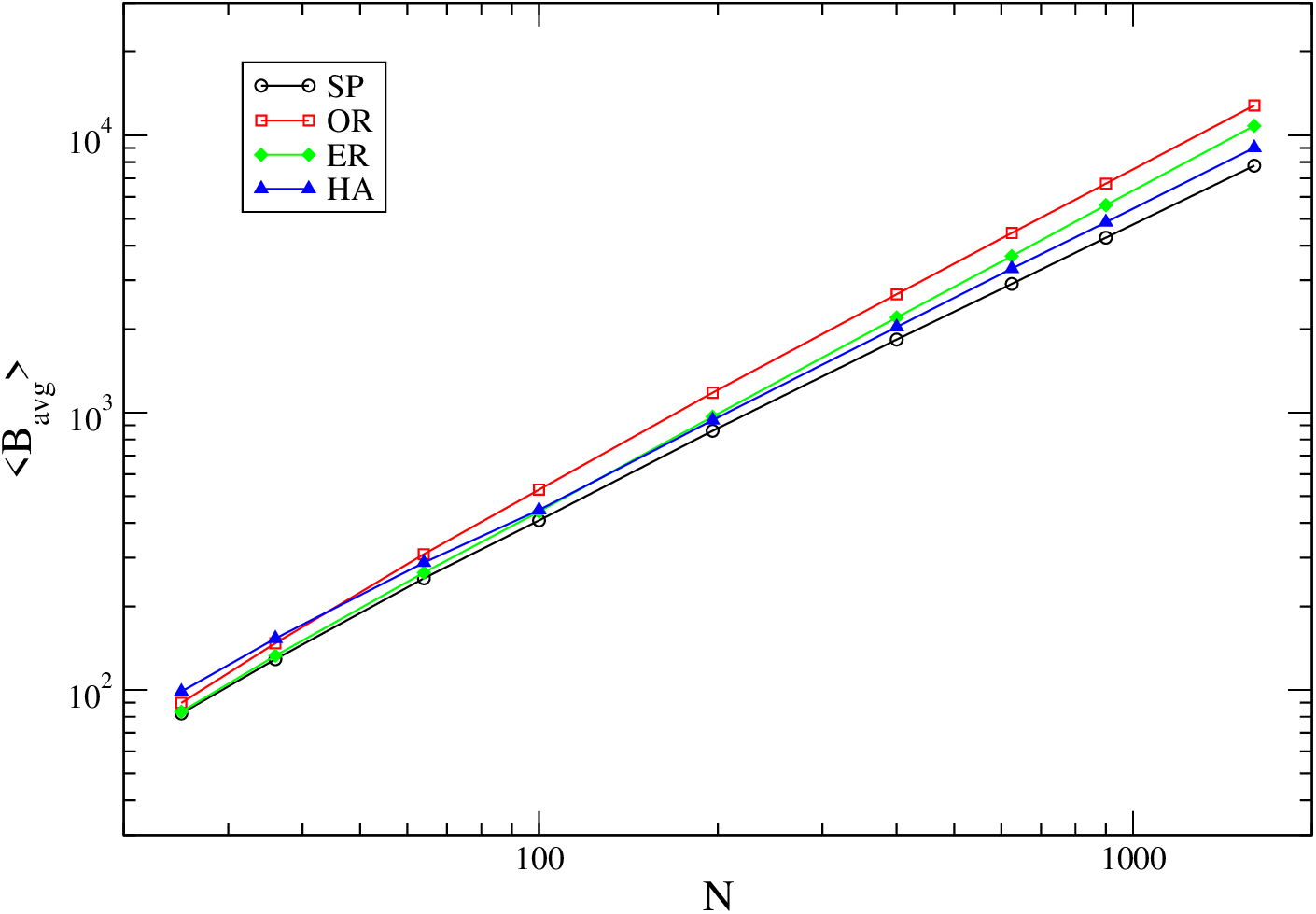}}
	\caption{(Color online) Ensemble average of the network average betweenness as a function of the number of nodes for four routing protocols.}
\end{figure}

Before presenting the results, we note that the average betweenness $B_{avg}$ on a given network using a given routing table provides an absolute lower bound for the maximum betweenness. This lower bound would be achievable only if one could optimize routing to the point where all nodes experience the same traffic. The difference between the maximum and average betweenness is also a measure of the width of the betweenness distribution. Moreover, the average betweenness in the case of shortest path routing with all weights set to 1 (which in the remainder of this Letter will be called shortest path routing) constitutes the lower bound for the average betweenness computed using any arbitrary routing table, since any changes from the shortest path routing (including those resulting from an optimization algorithm) will result in longer paths, thus adding to the sum of all betweennesses on the network. It is thus apparent that a good optimization algorithm is required to have at least two properties: (1) minimize the difference between the maximum and average betweenness, and (2) do this while keeping the difference between the average betweenness computed using the optimized routing table and the one computed using the shortest path routing table as low as possible.

In the following, networks of a given size $N$ are characterized by the ensemble averages of the maximum betweenness $<B_{max}>$ and average betweenness $<B_{avg}>$ computed over a set of 100 realizations of the network. Computer simulation results presented in Ref.\ \cite{Sreenivasan_arXiv} suggest a power-law functional dependence of $<B_{max}>$ on the network size $N$. Our results confirm this power-law dependence for $<B_{max}>$ and show a similar dependence in the case of $<B_{avg}>$.
\begin{figure}
	\scalebox{0.32}{\includegraphics{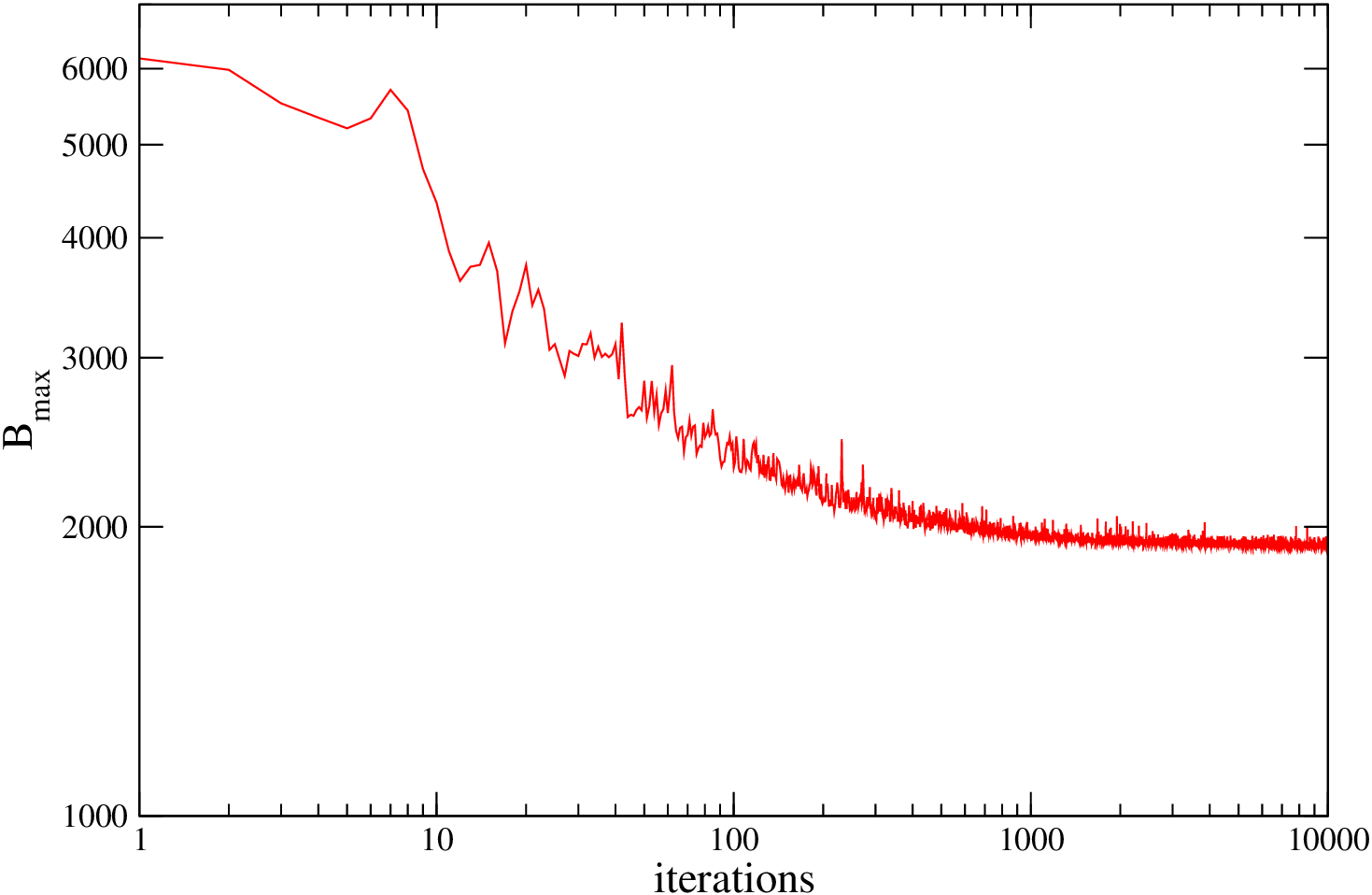}}
	\caption{Maximum betweenness as a function of the number of iterations for a network with 196 nodes.}
\end{figure}
\begin{figure}
	\scalebox{0.4}{\includegraphics{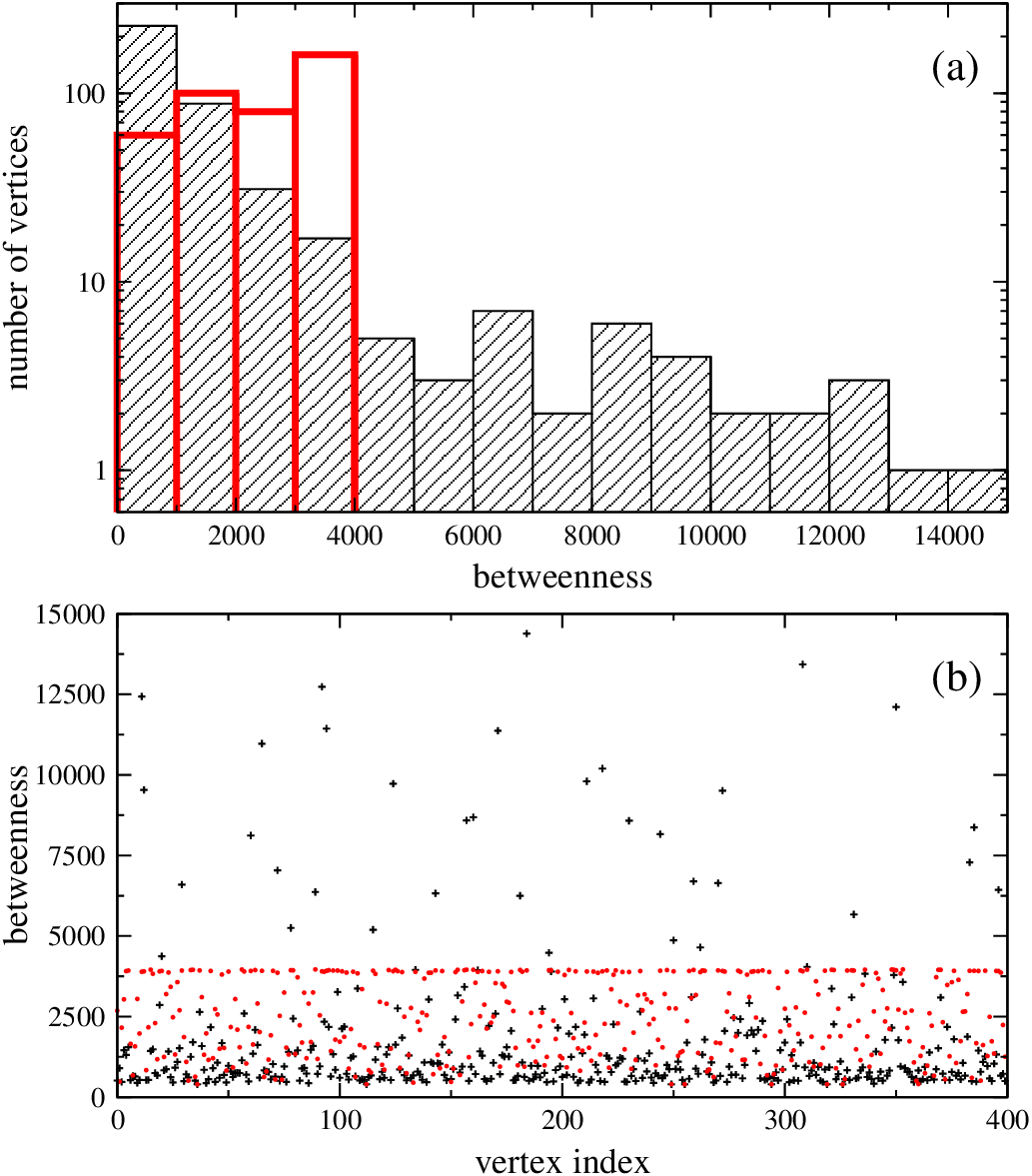}}
	\caption{(Color online) Distribution of node betweennesses before (black shaded bins in (a) and black crosses in (b)) and after (red hollow bins in (a) and red circles in (b)) optimization for a network with 400 nodes.}
\end{figure}

Results for the ensemble average of the maximum betweenness $<B_{max}>$ as a function of the network size $N$ for four routing protocols are presented in Fig.\ 1. Similar results for the ensemble average of the network average betweenness $<B_{avg}>$ are shown in Fig.\ 2. The results for the shortest path (SP) routing are represented by circles, while those obtained for the routing given by our optimization algorithm (OR) are represented by squares. For comparison, we also show results obtained using the efficient routing (ER) and hub avoidance (HA) protocols described in Refs.\ \cite{YanPRE} and \cite{Sreenivasan_arXiv}, represented by diamonds and triangles respectively. The exponents resulting from fitting the data points in each set are given in Table 1, where the quoted errors are $2\sigma$ estimates.
\begin{table*}
\begin{center}
	\begin{tabular}{|c||c||c||c||c|}
	\hline
		  & SP & OR & ER & HA \\
	\hline
      $<B_{avg}>$ & $1.088\pm 0.019$ & $1.186\pm 0.024$ & $1.165\pm 0.009$ & $1.080\pm 0.009$ \\
      $<B_{max}>$ & $1.634\pm 0.010$ & $1.185\pm 0.009$ & $1.315\pm 0.017$ & $1.542\pm 0.010$ \\
	\hline
	\end{tabular}
	\caption{Exponents of the $<B_{avg}>$ and $<B_{max}>$ power-law scaling with network size $N$.}
\end{center}
\end{table*}

It is apparent from Figs.\ 1 and 2 that the routing based on our optimization algorithm leads to a far smaller difference between the maximum and the average betweenness than in the case of the other three routing protocols. Moreover, the maximum and the average betweenness scale with $N$ with the same exponent, which is a strong argument in favor of the optimality of the routing. Finally, the difference $<B_{avg}>_{OR} - <B_{avg}>_{SP}$, while larger than $<B_{avg}>_{ER} - <B_{avg}>_{SP}$ or $<B_{avg}>_{HA} - <B_{avg}>_{SP}$ (which is explained by the need to have slightly longer paths around the hubs), is kept quite low and $<B_{avg}>_{OR}$ scales with the network size $N$ with an exponent only slightly higher than $<B_{avg}>_{SP}$.

As expected with a heuristic algorithm, the evolution of the maximum betweenness as a function of the number of iterations is not monotonic, but exhibits a decreasing trend and eventually the maximum betweenness ``converges" in the sense that it becomes confined to a narrow band. This is exemplified in Fig.\ 3, which is a plot of $B_{max}$ versus the number of iterations for a given network with 196 nodes. The algorithm can be implemented to pick the configuration corresponding to the global minimum of $B_{max}$.

Fig.\ 4 provides insight into how the algorithm works by comparing the initial and final betweenness distributions in the case of a network with 400 nodes. Fig.\ 4a shows histograms of the betweenness distribution before and after optimization, while Fig.\ 4b shows the betweennesses plotted against vertex index. Initially, the majority of the nodes have very low betweenness, but a small number of them are spread over a very wide betweenness range. After optimization, all node betweennesses are confined to a narrow band, whose upper edge is quite well defined. Most of them are uniformly distributed within the band, but there is a very sharp peak at the upper edge. There is a significant decrease in the number of very low betweenness nodes. A plot of the final (OR) betweenness versus the initial (SP) betweenness in the case of a network with 400 nodes is shown in Fig.\ 5. It is apparent from this plot that the algorithm performs remarkably well, by lowering the traffic through all nodes whose initial (shortest path) betweenness lies above a certain critical value until they all reach essentially the same critical betweenness. On the other hand, virtually all nodes whose initial betweenness lies below the critical value experience higher traffic, many of them (especially those with higher initial betweenness) reaching the critical value. It is still an open question whether an improved algorithm can achieve a lower critical betweenness by further raising the traffic through some initially low betweenness nodes. On the other hand, it is clear that not all low betweenness nodes can have their betweenness increased without unduly lengthening paths or increasing traffic through other nodes which are prone to congestion. The simplest examples are those of a small subnetwork whose only connection to the rest of the network is through a single link to a high degree (or otherwise high SP betweenness) node, or a triangle connected to the rest of the network only by containing such a node. In these cases, there is no way of diverting any of the high SP betweenness node's traffic between other nodes through the structures mentioned above. The latter will have low betweenness even in the case of rigorously optimal routing.
\begin{figure}
	\scalebox{0.32}{\includegraphics{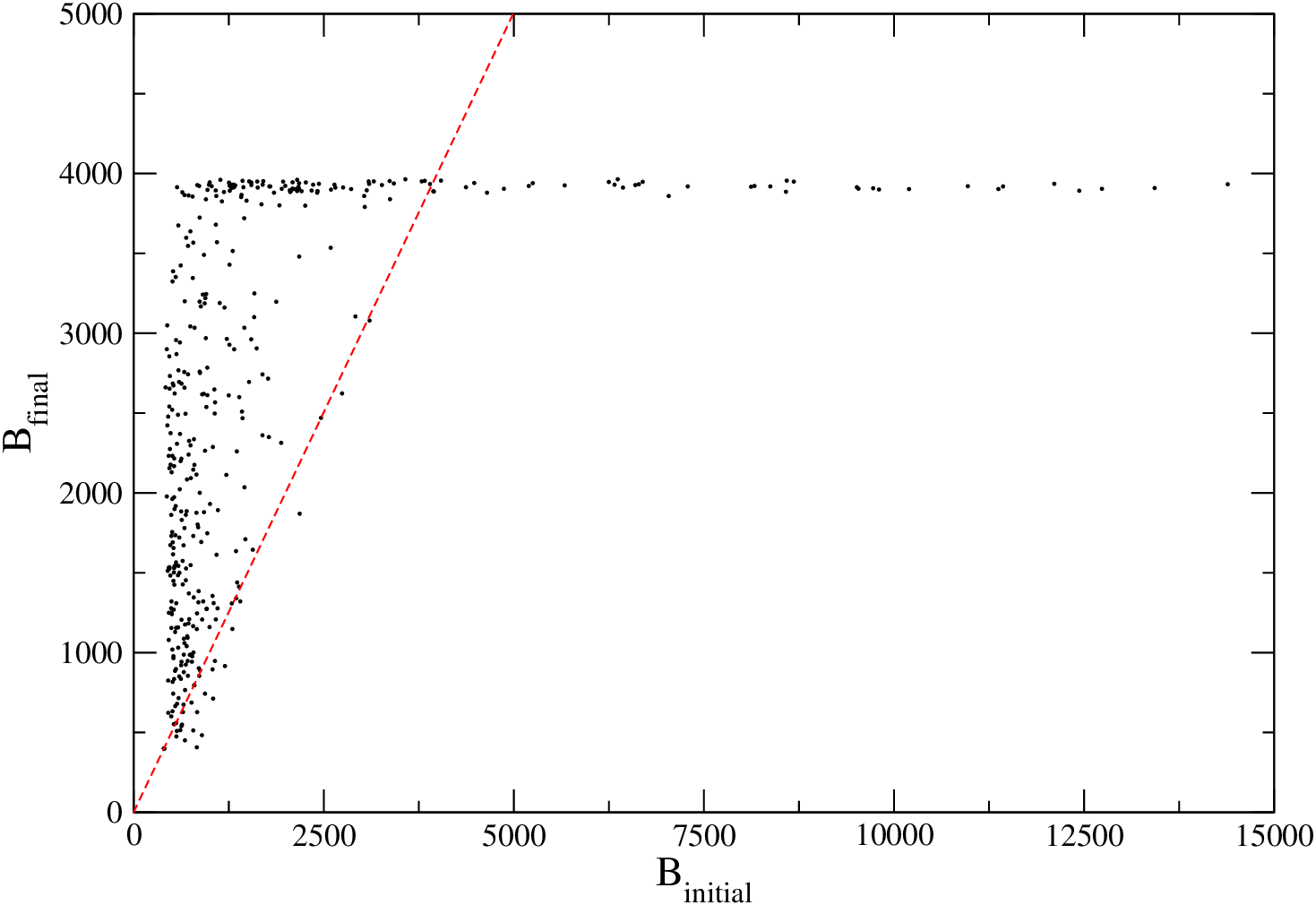}}
	\caption{Correlation plot of the final (OR) versus initial (SP) betweenness for a network with 400 nodes.}
\end{figure}

The difference between the SP and OR distribution of the travel times between the various nodes in situations when the SP routing does not lead to jamming is still an open question and will be the subject of a future study. However, we argue that, at least in situations of congested traffic, most travel times are shorter in the case of OR routing. It is known from queuing theory that the average queue length $<q>$ is given (assuming unity processing power) by \cite{Guimera,Allen}
\begin{equation}
	<q>=\frac{<w>}{1-<w>},
\end{equation}
\noindent with $<w>$ defined above Eq.\ 1. Due to this strongly nonlinear relationship, which diverges as $<w>$ approaches unity, it seems reasonable to assume that by avoiding the passage through hubs with very high betweenness most travel times become shorter, in spite of the fact that the routes pass through more nodes. This conclusion is also supported by the results for the average path length and average travel time presented in Ref.\ \cite{YanPRE}.

In summary, we have presented a simple heuristic algorithm for routing optimization on complex networks and demonstrated its usefulness for scale-free networks. This algorithm is useful in situations when network jamming is primarily due to queuing overload. Our results show that the application of this algorithm allows a network to bear significantly higher traffic than in the case of shortest path routing. Network capacity is improved by a factor which increases with network size according to a power law.

The authors acknowledge support from the NSF through grant No.\ DMR-0427538 and also from SI International through A.\ Williams of the Air Force Research Laboratory Information Directorate under contract No.\ FA8750-04-C-0258.


\end{document}